\documentclass[aps,pre,twocolumn,superscriptaddress]{revtex4-2}

\usepackage[utf8]{inputenc}
\usepackage{amsfonts}
\usepackage{amsmath}
\usepackage{amssymb}
\usepackage{graphicx}
\usepackage[colorlinks=true, allcolors=blue]{hyperref}

\newcommand{\ssdw}{SSD }
\newcommand{\ssdwnospace}{SSD}

\begin{document}

\title{Phase-field Modelling of Anisotropic Solid-State Dewetting on Patterned Substrates}

\author{Emma Radice}
\affiliation{L-NESS and Dept. of Materials Science, University of Milano-Bicocca, Milano, Italy}
\affiliation{Institute of Scientific Computing, TU Dresden, Dresden, Germany}

\author{Marco Salvalaglio}
\affiliation{Institute of Scientific Computing, TU Dresden, Dresden, Germany}
\affiliation{Dresden Center for Computational Materials Science, TU Dresden, Germany}

\author{Roberto Bergamaschini}
\affiliation{L-NESS and Dept. of Materials Science, University of Milano-Bicocca, Milano, Italy}
\email{roberto.bergamaschini@unimib.it}\makeatletter
\def\maketitle{
\@author@finish
\title@column\titleblock@produce
\suppressfloats[t]}
\makeatother
10.48550/arXiv.2405.03049

    \begin{abstract}
        We present a phase-field model for simulating the solid-state dewetting of anisotropic crystalline films on non-planar substrates. This model exploits two order parameters to trace implicitly the crystal free surface and the substrate profile in both two and three dimensions. First, we validate the model by comparing numerical simulation results for planar substrates with those obtained by a conventional phase-field approach and by assessing the convergence toward the equilibrium shape predicted by the Winterbottom construction. We then explore non-planar geometries, examining the combined effects of surface-energy anisotropies and parameters controlling the contact angle. Our findings reveal that crystalline particles on curved supports lose self-similarity and exhibit a volume-dependent apparent contact angle, with opposite trends for convex versus concave profiles. Additionally, we investigate the migration of faceted particles on substrates with variable curvature. Applying this model to experimentally relevant cases like spheroidal and pit-patterned substrates demonstrates various behaviors that could be leveraged to direct self-assembly of nanostructures, from ordered nanoparticles to interconnected networks with complex topology.
    \end{abstract}

  \maketitle

\section{Introduction}
Solid-state dewetting (\ssdwnospace) refers to the breakup and rearrangement of a thin crystalline film on a weakly bonded foreign substrate through surface diffusion. Mainly driven by the minimization of the surface energy and activated at relatively high temperatures, this process involves the evolution of thin films toward structures with lower surface-to-volume ratios, generally corresponding to three-dimensional (3D) islands \cite{ThompsonARMR2012}. Several solid materials of broad technological interest, like semiconductors (e.g., Si, Ge) or metals (e.g., Ni, Co, Au), have been shown to dewet on foreign substrates; see Refs.~\cite{ThompsonARMR2012, RuffinoPSSA2015, LeroySSR2016} and references therein. Although detrimental for applications based on planar thin films, \ssdw provides a kinetic pathway from a simple two-dimensional (2D) layer to various complex morphologies, including pierced films, nanowires, and three-dimensional agglomerates. It may then be exploited for nanostructure self-assembly with applications ranging from optoelectronics \cite{AbbarchiACSNANO2014, BollaniNC2019} to catalysis \cite{AltomareCS2016}.

Typically, \ssdw initiates at the edges of finite-size films or through the spontaneous emergence of voids within the film. It results in disordered arrangements of islands. However, several strategies have been proposed to control the self-assembly of nanostructures by \ssdw and enforce ordering \cite{RuffinoPSSA2015,LeroySSR2016}. Among them, the exploitation of film patterning techniques by top-down lithography before annealing offered a high degree of control and customization \cite{YeADVMAT2011, SundarML2012, BenkouiderNT2015, NaffoutiSCIADV2017, BollaniNC2019}. An alternative approach that enforces ordering by triggering dewetting at selected locations consists of modulating the substrate topography, e.g., into ripples \cite{PetersenJAP2008,WangJMS2012,LuSR2016}, pits \cite{OhSMALL2009,WangJMSME2011,GiermannJAP2011}, columns \cite{WallaceNANOSADV2020}, spheres \cite{YangADVFUNCTMAT2011,KrupinskiNANOT2015}.

Microscopically, \ssdw stems from the stochastic diffusion of adatoms at the surface of the crystalline film. A few efforts have been made to model this dynamics at the atomic scale by Kinetic Monte-Carlo \cite{PierreLouisPRL2007,SaitoCRP2013,TrautmannAPL2017}. However, to approach the large scale and long times of typical experiments, a coarser description based on continuum models is needed. The earliest models applicable to \ssdw can be traced back to the seminal work on surface diffusion by Mullins \cite{MullinsJAP1957}, where, under the assumption of isotropic surface energy, the evolution of a surface profile follows the local gradients of its mean curvature. The film-substrate interaction is introduced as a boundary condition (BC) enforcing a specific contact angle $\theta$ at the triple junction. This is defined by the well-known Young equation \cite{ThompsonARMR2012} $\gamma\cos\theta=\sigma$, where $\gamma$ is the surface energy density of the film material and $\sigma=\gamma_{\rm s}-\gamma_i$ is the difference in energy density between the substrate surface $\gamma_{\rm s}$ and the film-substrate interface $\gamma_{\rm i}$; see, e.g., Refs.~\cite{WongAM2000,CheynisCRP2013}.

One peculiar feature of crystalline materials is that free surfaces exhibit preferential orientations, thus leading to faceted morphologies. The equilibrium geometry of a supported crystalline particle can be determined through the Winterbottom construction \cite{WinterbottomAM1967, BaoSIAMJAM2017} by considering its anisotropic surface energy density $\gamma(\theta,\phi)$, with $\theta$ and $\phi$ the polar and azimuthal angles in spherical coordinates, and the wetting parameter $\sigma$ above. More precisely, in the presence of anisotropic surface energies, the contact-angle BC modifies into the Herring-Young equation \cite{JiangSIAMJAM2020}:
\begin{equation}\label{eq::herringyoung}
    \gamma(\theta,\phi)\cos\theta-\frac{\partial\gamma(\theta,\phi)}{\partial\theta}\sin\theta=\sigma,
\end{equation}
with $\theta$ and $\phi$ evaluated along the contact line. Note that these angles may vary on the substrate. Then, the $\theta$ values satisfying eq.~\eqref{eq::herringyoung} depend on $\phi$ and generally deviate from the isotropic value as a function of the anisotropy strength. Moreover, in the case of non-convex $\gamma(\theta,\phi)$ functions, i.e., in the so-called strong anisotropy regime, eq.~\eqref{eq::herringyoung} can admit multiple solutions \cite{BaoSIAMJAM2017}. A few models \cite{DornelPRB2006,ZuckerCRP2013,WangPRB2015,JiangSM2016,BaoJCP2017} have been proposed to model the evolution of a faceted profile during \ssdwnospace. Recently, in Refs.~\cite{JiangPRM2018,ZhaoAM2024}, nonplanar substrates were also considered. However, these so-called sharp-interface approaches were limited to 2D. Their extension to a fully 3D description, even if possible \cite{JiangSIAMJAM2020}, requires significant implementation efforts to retain generality.

Phase-field (PF) models \cite{LiCCP2009} have emerged as suitable approaches to implicitly describe complex 3D morphologies while naturally dealing with topological changes, such as the breakup and coalescence of crystalline domains. Several studies in the recent literature took profit of such method for simulating \ssdw in 3D, yet considering isotropic $\gamma$ \cite{JiangACTAM2012, NaffoutiSCIADV2017,BackofenIJNAM2019,VermaJAC2020}, even tackling the case of curved substrates by multiphase approaches \cite{BretinESAIM2023,ShiDCDS2024}. In Ref.~\cite{DziwnikNLIN2017}, a PF model including $\gamma$ anisotropy and a natural BC for the film-substrate contact interaction was demonstrated to asymptotically converge to the sharp-interface model satisfying the Herring-Young condition (eq.~\eqref{eq::herringyoung}), yet for the 2D case only. An extension to 3D has been recently discussed in Ref.~\cite{GarckeJNS2023}, and a few simulation studies of \ssdw with anisotropic $\gamma$ have been reported \cite{BollaniNC2019,GranchiOE2023}, along with an affine approach based on the Level Set method \cite{GavhaleAM2022,EtoileAM2024}. Nonetheless, neither of these approaches provides a comprehensive 3D modelling of \ssdw for crystalline materials, i.e., including $\gamma$ anisotropies and compatible contact angles, suitable to tackle the complexity of a broad range of experimental settings \cite{PetersenJAP2008,OhSMALL2009,WangJMSME2011,GiermannJAP2011,WallaceNANOSADV2020,KrupinskiNANOT2015} involving both planar and curved substrates.

In this work, we develop a framework based on the PF model to simulate anisotropic \ssdw in 3D on substrates of arbitrary geometry. To this goal, we consider a PF model for free-standing geometries \cite{torabiPRSA2009,SalvalaglioAMMS2021b} and couple it with the implicit description of the substrate leveraging the extended smoothed boundary method reported in Ref.~\cite{YuMSMSE2012}. For the sake of simplicity, we focus on strictly convex $\gamma$ functions, i.e., on weakly anisotropic regimes. This choice allows us to focus on the simplest model for anisotropic surface diffusion, technically neglecting higher order regularizations \cite{torabiPRSA2009} while inspecting and showcasing the newly introduced coupling with the substrate. The proposed framework is shown to recover known results concerning planar substrates and isotropic dewetting on curved surfaces. Moreover, we show how it enables simulations of anisotropic \ssdw on nontrivial substrate geometry and discuss implications towards understanding and controlling such complex scenarios.

The paper is organized as follows. In Sect.~\ref{sec::methods}, we describe the anisotropic PF model for tracing the surface evolution and its confinement on top of a substrate region, along with the appropriate BC. Then, in Sect.~\ref{sec::results} we report simulation results. First, in Sect.~\ref{sec::results::proof}, we test the reliability of the model for simple crystal structures on a planar substrate, comparing the profiles with the ones from a PF model without the diffuse-boundary description of the substrate and with sharp-interface theoretical predictions. Then, we inspect the dewetting dynamics of faceted crystals on non-planar substrates, starting from spherical surfaces in Sect.~\ref{sec::results::spherical} and then moving to variable curvature geometries in Sect.~\ref{sec::results::variablek}. Last, in Sect.~\ref{sec::results::pits} we focus on the technology-relevant case of dewetting on a pit-patterned substrate. The main conclusions are summarized in Sect.~\ref{sec:conclusions}.

\section{Model}\label{sec::methods}
We consider a PF model describing a crystalline phase in a domain $\Omega$ by means of a smooth order parameter $\varphi$, varying from 1 in the crystal to 0 in the surrounding vacuum \cite{LiCCP2009}. The surface energy of the crystal is approximated by the functional \cite{torabiPRSA2009,SalvalaglioAMMS2021b}
\begin{equation}\label{eq::free-energy}
    F=\int \frac{\gamma({\bf \hat{n}})}{g(\varphi)} \left(\frac{\epsilon}{2}|\nabla\varphi|^2 + \frac{1}{\epsilon}W(\varphi)\right) d{\bf x},
\end{equation}
with $\epsilon$ the width of the diffuse interface at the crystal surface (conventionally identified as the $\varphi=0.5$ isoline), $\gamma({ \hat{\bf n}})$ the anisotropic surface energy density as a function of the surface normal direction ${\hat{\bf n}}=-\nabla\varphi/|\nabla\varphi|$, $W(\varphi)=18\varphi^2(1-\varphi)^2$ a double-well potential, and $g(\varphi)=6|\varphi||1-\varphi|$~\footnote{$g(\varphi)$ is made non-singular by adding a small regularization parameter $\delta\sim10^{-6}$: $g(\varphi)=6|\varphi||1-\varphi| +\delta$.} a stabilizing function allowing for an improved accuracy in the description of surface dynamics \cite{SalvalaglioDDCHiso2021,SalvalaglioAMMS2021b}.

The evolution of $\varphi$ reproducing surface diffusion is described by the Cahn-Hilliard equation:
\begin{equation}\label{eq::cahn-hilliard0}
    \frac{\partial\varphi}{\partial t}= \nabla\cdot (M(\varphi) \nabla\mu),
\end{equation}
with $M=M_0(36/\epsilon)\varphi^2(1-\varphi)^2$ a degenerate mobility, scaled by the factor $M_0$. $\mu=\delta F/\delta\varphi$ corresponds to the surface chemical potential with $F$ from eq.~\eqref{eq::free-energy}. Exploiting the approximation $(1/\epsilon)W(\varphi)\,\approx\,(\epsilon/2)|\nabla\varphi|^2$ (exact in the asymptotic limit $\epsilon\rightarrow 0$ \cite{SalvalaglioDDCHiso2021}) it reads
\begin{equation}\label{eq::mu0}
    g(\varphi) \mu = -\epsilon \nabla \cdot (\gamma \nabla\varphi) + \frac{\gamma}{\epsilon}W'(\varphi)+\epsilon\nabla\cdot\left(|\nabla\varphi|{\bf P}\nabla_{\hat{\bf n}}\gamma\right),
\end{equation}
where ${\bf P}={\mathbf{I}-\hat{\bf n}\otimes\hat{\bf n}}$. 
The equations \eqref{eq::cahn-hilliard0}-\eqref{eq::mu0}, supplemented by zero-flux Neumann BCs on $\partial \Omega$, give a full description of the surface diffusion dynamics, including surface faceting of free-standing crystals in the weak-anisotropy regime. We remark that the addition of a Willmore energy regularization term would be needed to tackle strong-anisotropy \cite{LiCCP2009, SalvalaglioCGD2015}, which is however beyond the scope of the present work. 

If the crystal (i.e., the $\varphi=1$ region) intersects a portion of the domain boundary $\Gamma_w\subseteq\partial\Omega$, the energetics underlying the interaction with the substrate can be accounted for by adding a wetting energy integral \cite{YeADVMAT2011} to eq.~\eqref{eq::free-energy}
\begin{equation}\label{eq:bintegral}
 \int_{\Gamma_w} f_w(\varphi) d{\bf x}=\int_{\Gamma_w} \left[\gamma_{\rm s}-\sigma\varphi^2 (3-2\varphi)\right]d{\bf x},
\end{equation} 
where $f_w(\varphi)$ is an interpolation function between the surface free-energy density of the substrate, $\gamma_{\rm s}$, and the substrate-crystal interface energy density, $\gamma_{\rm i}$, and $\sigma=\gamma_{\rm s}-\gamma_{\rm i}=\gamma\cos{\theta_{\rm iso}}$, with $\theta_{\rm iso}$ the equivalent contact angle for isotropic $\gamma$.

The first variation of the boundary integral \eqref{eq:bintegral} then introduces the Neumann BC:
\begin{equation}\label{eq::herringyoung_pf}
    \epsilon \left[ \gamma\nabla\varphi - |\nabla\varphi|{\bf P}\nabla_{\hat{\bf n}}\gamma \right]\cdot{\hat{\bf n}_w} = g(\varphi)\sigma,
\end{equation}
where ${\hat{\bf n}_w}$ is the (inward) normal to the $\Gamma_w$ boundary and $g(\varphi)\sigma=f_w'(\varphi)$. It has been shown \cite{DziwnikNLIN2017} that, in the $\epsilon\rightarrow0$ limit, such condition converges to the Young-Herring equation \eqref{eq::herringyoung} (or to the simpler Young equation $\gamma\cos\theta=\sigma$ for isotropic $\gamma$). Previous studies showed how the numerical integration of eqs.~\eqref{eq::cahn-hilliard0}-\eqref{eq::mu0} with the BC \eqref{eq::herringyoung_pf} returns the expected anisotropic shapes with the appropriate contact-angles on a planar boundary \cite{DziwnikNLIN2017, GarckeJNS2023}. 

\begin{figure}
  \includegraphics[width=\columnwidth]{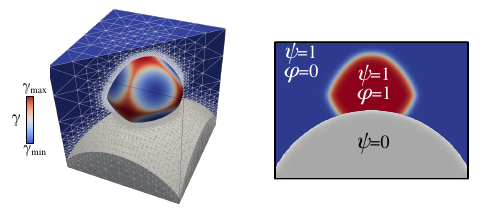}
  \caption{Simulation domain for a 3D faceted particle on a curved substrate (left) and illustration of the two phase-fields $\varphi$ and $\psi$ in the diagonal cross-section (right). The substrate region (gray) is identified by $\psi\le0.5$, while the crystalline particle (colored according to the local surface energy $\gamma$ in the 3D view) is the $\varphi=0.5$ contour within the $\psi>0.5$ region. The remaining integration domain (blue) corresponds to the vacuum. The $\varphi$ field, although defined in the whole domain, has no physical meaning within the substrate region. The adaptive mesh used for numerical simulations, which is finer at the crystal and substrate surfaces, is also illustrated.}\label{fig::fig1}
\end{figure}

The approach above can be applied to simulate \ssdw over a substrate with arbitrary geometries by defining $\partial \Omega$ accordingly. This, however, means solving eqs.~\eqref{eq::cahn-hilliard0}-\eqref{eq::mu0} in domains with complex shapes, in general requiring ad-hoc implementations. To achieve more flexibility in the definition of arbitrary substrate geometries, and decouple its definition from $\Omega$, we here exploit an alternative formulation \cite{SchellingerhoutADVFUN2023} by tracing the substrate profile implicitly via a second (static) order parameter $\psi$ defined as
\begin{equation}
    \psi({\bf x})=\frac{1}{2}\left[\tanh\left(\frac{3 d({\bf x})}{\epsilon}\right)-1\right],
\end{equation}
with $d({\bf x})$ the signed-distance of each point ${\bf x}$ in the domain from the nominal substrate surface. $\epsilon$ is chosen to be equal to the interface width of $\varphi$ (imposed via parametrization of \eqref{eq::free-energy}). Then, $\psi$ is set to 0 within the substrate and 1 in the outer region, where the crystal is defined. A schematic of the simulation domain illustrating the considered phase fields is reported in Figure~\ref{fig::fig1}.

Following Ref.~\cite{YuMSMSE2012}, the evolution equations \eqref{eq::cahn-hilliard0}-\eqref{eq::mu0} for the crystal phase $\varphi$ can then be confined in the region external to the substrate by multiplying both the left- and right-hand side of eqs.~\eqref{eq::cahn-hilliard0}-\eqref{eq::mu0} by $\psi$:
\begin{equation}
\begin{split}
    \psi\frac{\partial\varphi}{\partial t} =& \nabla\cdot (\psi M(\varphi) \nabla\mu) - M(\varphi)\nabla\mu\cdot\nabla\psi,  \\
     \psi g(\varphi)\mu = & -\epsilon \nabla \cdot (\psi \gamma \nabla\varphi) + \psi\frac{\gamma}{\epsilon}W'(\varphi)+\\
    & +\epsilon\nabla\cdot\left(\psi|\nabla\varphi|{\bf P}\nabla_{\hat{\bf n}}\gamma\right) + \\
    & + \epsilon \left[ \gamma\nabla\varphi - |\nabla\varphi|{\bf P}\nabla_{\hat{\bf n}}\gamma  \right]\cdot\nabla\psi.
    \label{eq::sbm1}
\end{split}
\end{equation}
By recalling that ${\hat{\bf n}_\psi}=\nabla\psi/|\nabla\psi|$ is the (inward) normal unit vector at the substrate surface, the last terms of eqs.~\eqref{eq::sbm1} can be expressed via the actual BC at the substrate surface. More precisely, the term $M\nabla\mu\cdot\nabla\psi = |\nabla\psi|{\bf J}\cdot {\hat{\bf n}_\psi}=0$ corresponds to the flux of $\varphi$ (i.e., the flux of material) to the substrate, which is null for the required volume conserving dynamics. The term $\epsilon \left[ \gamma\nabla\varphi - |\nabla\varphi|{\bf P}\nabla_{\hat{\bf n}}\gamma \right]\cdot\nabla\psi$ corresponds to the mapping of the Neumann BC in \eqref{eq::herringyoung_pf} along the substrate surface. Then,
\begin{equation}\label{eq::herringyoung_pf_psi}
\begin{split}
    \epsilon \left[ \gamma\nabla\varphi - |\nabla\varphi|{\bf P}\nabla_{\hat{\bf n}}\gamma  \right] \cdot\nabla\psi &= |\nabla\psi| g(\varphi)\sigma \\
    &\approx\frac{1}{\epsilon}g(\psi)g(\varphi)\sigma,
\end{split}
\end{equation}
where the approximation $\epsilon|\nabla\psi|\approx g(\psi)$ (exact in the limit $\epsilon \rightarrow 0$) was considered. Finally, the evolution equations for the dynamics of the anisotropic surface profile, denoted by $\varphi$, on the substrate traced by $\psi$ with prescribed contact condition set by the value of $\sigma$, results in
\begin{equation}
\begin{split}
    \psi\frac{\partial\varphi}{\partial t} =& \nabla\cdot (\psi M(\varphi) \nabla\mu),  \\
     \psi g(\varphi)\mu = & -\epsilon \nabla \cdot (\psi \gamma \nabla\varphi) + \psi\frac{\gamma}{\epsilon}W'(\varphi)+\\
    & +\epsilon\nabla\cdot\left(\psi|\nabla\varphi|{\bf P}\nabla_{\hat{\bf n}}\gamma\right) + \\
    & + \frac{1}{\epsilon}g(\psi)g(\varphi)\sigma.
\label{eq::sbm}
\end{split}
\end{equation}
For the sake of simplicity, it is assumed that the crystal phase ($\varphi$=1) does not intersect the integration cell boundaries so that zero-flux Neumann BCs for $\varphi$ can be set over $\partial \Omega$. This approach allows for inspecting arbitrarily designed substrate geometries using the field $\psi$ inside simple integration domains $\Omega$.

\subsection{Numerical details}
Following most of previous literature studies, for simple tests in 2D a cosine function is used to define the $\gamma$ anisotropy as 
\begin{equation}\label{eq::cosgamma}
 \gamma(\vartheta)=1+\alpha\cos{\left[N(\vartheta-\phi)\right]},
\end{equation}
with $\vartheta$ the polar angle, $\alpha$ the anisotropy strength, $N$ the number of equivalent facets (i.e. the $\gamma$ mimina) along the circle and $\phi$ a phase factor allowing to rotate the axis. For consistency, in the present study we only report results for a square anisotropy ($N=4$) and set $\alpha=0.06$. For this value, $\gamma(\vartheta)$ is convex but close to the convex-concave transition (see Fig.~\ref{fig::fig2}).

For 3D simulations, instead, we exploit the convenient formulation of $\gamma({\hat{\bf n}})$ from Ref.~\cite{SalvalaglioCGD2015}, so to consider energy minima corresponding to crystallographic planes with orientation ${\hat{\bf m}}$ independently,
\begin{equation}\label{eq::msgammma}
    \gamma({\hat{\bf n}})=\gamma_0\left[1-\sum_i \alpha_i({\hat{\bf n}}\cdot{\hat{\bf m}_i})^w\Theta({\hat{\bf n}}\cdot{\hat{\bf m}_i})\right],
\end{equation}
where $\gamma_0$ is a scaling parameter returning the value of surface energy at unfaceted orientations, $\alpha_i$ is the relative depth of the $i$-th $\gamma$ minimum along ${\hat{\bf m}_i}$ direction, $w$ controlling its width, and $\Theta$ is the Heaviside function allowing to consider geometries without inversion-symmetry.

Given the self-similarity of the dynamics with respect to any rescaling of the spatial coordinates, we take $\epsilon$ as the unit of length. Additionally, time is expressed in arbitrary units, proportional to the mobility parameter $M_0$ (with $\gamma_0$ absorbed as a simple multiplicative factor). We remark that the numerical resolution on the onset of the film breakup occurring during \ssdw inherently depends on $\epsilon$ in diffuse domain approaches. To allow for comparisons between simulations featuring such a behavior, we set the same value of $\epsilon$ for all of them.

The numerical integration is performed by using the AMDiS Finite-Element Method toolbox \cite{VeyCVS2007, WitkowskiACM2015}, exploiting a semi-implicit time-integration scheme using adaptive time stepping ensuring decreasing energy and numerical convergence. Adaptive mesh refinement is also exploited (see Fig.~\ref{fig::fig1}) with maximum resolution $\sim 0.1 \epsilon$ at the $\varphi$=0.5 interface. Given that the $\psi$ field is analytically interpolated on the mesh at each adaptation step, a coarse mesh is used away from the contact point. A refined mesh for $\psi$ is only considered for plotting.

All simulation profiles reported in the following correspond to the $\varphi=0.5$ iso-surface, clipped within the $\psi<0.5$ domain.

\section{Results and discussion}\label{sec::results}

\subsection{Model testing on planar substrates} \label{sec::results::proof}

\begin{figure*}[tb!]
  \centering
  \includegraphics[width=\textwidth]{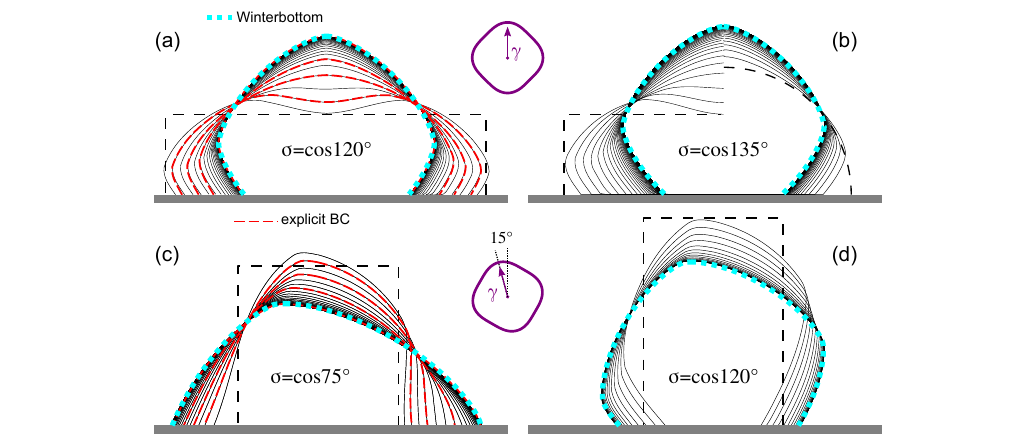}
  \caption{2D \ssdw simulations of a crystalline particle with the four-fold $\gamma$ anisotropy defined in eq.~\eqref{eq::cosgamma},, rotated as in the insets, on top of a planar substrate. The initial profile is shown by the black dashed line. The simulation sequence is reported at constant time intervals by the black (solid) contour lines. The final stages converge to the equilibrium shape, as obtained by the Winterbottom construction (cyan, dotted). The red, dashed profiles in panels (a) and (c) are obtained by solving the same evolution problem by implementing the explicit Neumann BC from eq.~\eqref{eq::herringyoung_pf} instead of the implicit definition of the substrate by $\psi$ (selected simulation times are chosen, showcasing an almost perfect match with corresponding black solid lines).}\label{fig::fig2}
\end{figure*}

We test the reliability and accuracy of the model illustrated in Sect.~\ref{sec::methods} using the well-known case of dewetting on a planar substrate as a benchmark. We illustrate several simulations for isotropic and anisotropic $\gamma$, checking the convergence of the evolution dynamics to the equilibrium profiles predicted by the Winterbottom construction. The dynamical behavior is validated by comparing the complete evolution sequences with the ones obtained by using the conventional PF model implementing the contact condition from eq.~\eqref{eq::herringyoung_pf} directly at the bottom of the simulation cell.

In Fig.~\ref{fig::fig2}, numerical simulations of \ssdw in 2D with the four-fold $\gamma$ anisotropy defined in eq.~\eqref{eq::cosgamma} are illustrated. Different wetting conditions are considered by varying the relative orientation $\phi$ between the crystal and the substrate and the contact-parameter $\sigma$. Fig.~\ref{fig::fig2}(a) illustrates the evolution sequence of an initially rectangular profile with $\sigma=\cos{120^{\circ}}$ (corresponding to an actual contact angle of $\approx128^\circ$ from solution of eq.~\eqref{eq::herringyoung}). The outcome of the proposed approach matches almost perfectly the simulation with explicit Neumann BC (eq.~\eqref{eq::cahn-hilliard0} and \eqref{eq::mu0} with BC~\eqref{eq::herringyoung_pf}) reported as dashed red lines at a few representative stages. Moreover, the final shape corresponds to the one predicted by the Winterbottom construction (cyan, dotted line). Another example is shown in Fig.~\ref{fig::fig2}(b), where the contact condition is changed to $\sigma=\cos{135^{\circ}}$ (actual contact angle $\approx133^\circ$) and two different initial profiles, a rectangle (left half) and a semicircle (right half), are considered. Both simulations converge to the same final profile, which once again corresponds to the theoretical equilibrium shape. We also inspect arbitrary rotations of the crystallographic axis with respect to the substrate normal, producing asymmetric equilibrium shapes with different contact angles at the left and right borders. Two cases are analyzed for a $15^{\circ}$ counterclockwise rotation of the $\gamma$ function used in Figs.~\ref{fig::fig2}(a) and \ref{fig::fig2}(b). In particular, in Fig.~\ref{fig::fig2}(c), a $\sigma=\cos{75^{\circ}}$ wetting condition is considered, corresponding to actual contact angles of $\approx82^\circ$ (on the left) and $\approx67^\circ$ (on the right). In Fig.~\ref{fig::fig2}(d), $\sigma=\cos{120^{\circ}}$ is set, resulting in contact angles of $\approx 121^\circ$  (on the left) and $\approx 133^\circ$  (on the right). These simulations also show convergence to asymmetric equilibrium (Winterbottom) shapes, with evolution pathways again consistent with the results obtained by classical PF modelling with appropriate Neumann BC (red dashed lines) as showcased in Fig.~\ref{fig::fig2}(c).

\begin{figure}[tb]
    \includegraphics[width=1\linewidth]{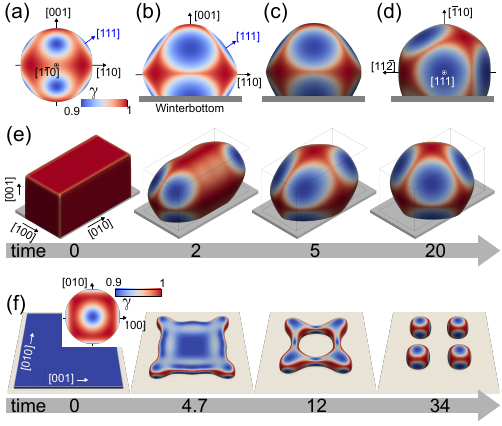}
  \caption{3D \ssdw simulations of faceted particles on a planar substrate, with contact parameter $\sigma=\cos{120^\circ}$.  (a) $\gamma$-plot of the \{111\} octahedral anisotropy used in (b-e). (b) Equilibrium shape computed by the Winterbottom construction with (001) vertical axis. (c) Final shape from the PF evolution from a hemispherical particle under the same conditions of (b), and (d) when rotating the vertical axis in the $[\bar{1}10]$ crystallographic direction. (e) PF evolution sequence of a ($2:1:1$)-cuboid particle tending to the same equilibrium shape in (b). (f) Dewetting simulation for a thin-film patch of 0.028 height-to-width aspect ratio (height equal to $2\epsilon$) with cubic $\gamma$ anisotropy reported in the $\gamma$-plot inset (\{100\} minima with $\alpha=0.1$ e $w=10$).}
  \label{fig::fig3}
\end{figure}

Simulations of 3D \ssdw on planar substrates are shown in Fig.~\ref{fig::fig3}. In Fig.~\ref{fig::fig3}(a)-(e) an octahedral $\gamma$ anisotropy, set via eq.~\eqref{eq::msgammma} with identical $\left\langle111\right\rangle$ minima ($\alpha=0.1$, $w=10$) is considered, as illustrated in Fig.~\ref{fig::fig3}(a). In Fig.~\ref{fig::fig3}(b) we report the theoretical Winterbottom shape for a crystal particle grown with the $[001]$ vertical axis and contact parameter $\sigma=\cos{120^{\circ}}$ while in Fig.~\ref{fig::fig3}(c) we show the final equilibrium shape obtained by a dewetting simulation from a hemispherical particle, quantitatively consistent with the Winterbottom shape in Fig.~\ref{fig::fig3}(b).
This holds true for any orientation of the crystal axis. See, for instance, the case of Fig.~\ref{fig::fig3}(d), where the vertical axis has been aligned with the $[\bar{1}10]$ direction. 

Equilibrium configurations are reached for different initial geometries, too. As an example, we report the evolution sequence for a \{100\} cuboid in Fig.~\ref{fig::fig3}(e). After a sudden local re-faceting of the shorter $yz$ edges, such to expose the lowest-energy \{111\} planes, a progressive retraction along the $\left\langle100\right\rangle$ direction drives the crystal shape toward the equilibrium shape as in Fig.~\ref{fig::fig3}(b). 

As a last example, we consider a \ssdw scenario leading to topological changes. In Fig.~\ref{fig::fig3}(f), we consider a low-aspect-ratio crystalline film patch, with a cubic $\gamma$ anisotropy set by $\left\langle100\right\rangle$ minima ($\alpha=0.1$, $w=10$) reported in the inset. As observed in previous studies \cite{YeADVMAT2011,NaffoutiSCIADV2017}, the material accumulates at the patch rims, especially at the corners, while being extracted from the central region until a hole opens at the center of the patch. At later stages, the rims further break with the formation of four particles, which converge to the expected equilibrium shape, i.e., a supported, smooth cube identical to the one predicted by the Winterbottom construction.

\subsection{Dewetting dynamics on spherical surfaces}\label{sec::results::spherical}
We here inspect the \ssdw dynamics on substrate profiles with a constant curvature, i.e., spherical surfaces, leveraging its implicit representation.

\begin{figure}[tb]
    \includegraphics[width=1\linewidth]{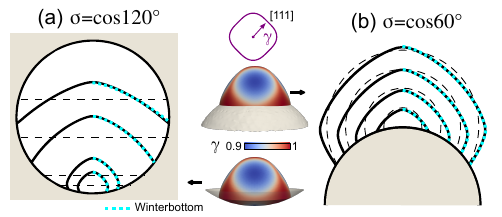}
  \caption{Particles on circular/spherical substrates. (a, b) 2D equilibrium profiles (solid lines) obtained as final states of PF \ssdw simulations from the initial shape traced by dashes, for crystals with the four-fold $\gamma$ anisotropy from eq.~\eqref{eq::cosgamma} on a concave (a) and (b) convex circular substrate, with $\sigma =\cos120^\circ$ and $\sigma=\cos60^\circ$ respectively. The cyan dotted lines report the corresponding effective Winterbottom shapes. The two 3D shapes in the inset show analogous final shapes obtained for octahedral $\gamma$ anisotropy as from Fig.~\ref{fig::fig3}(a) after \ssdw of initially hemispherical particles with a radius half of the underlying spherical substrates.}
  \label{fig::fig4}
\end{figure}

In Fig.~\ref{fig::fig4}, we consider either a concave (a) or convex (b) 2D circular substrate (gray region) and inspect the equilibrium shape of faceted particles with the square $\gamma$ anisotropy used in Fig.~\ref{fig::fig2}(a,b) for different sizes.
More precisely, in Fig.~\ref{fig::fig4}(a), the initial profiles are set as horizontal chords at different heights, and the wetting parameter is $\sigma=\cos{120^\circ}$. In Fig.~\ref{fig::fig4}(b), the initial particles are defined by circular sectors of different radii, centered at the summit of the substrate and $\sigma=\cos{60^\circ}$. The final profiles exhibit different shapes because of their different areas (volumes), which move the contact points at locations with different substrate orientations, thus breaking the self-similarity of the Winterbottom shape on planar substrates in the same way as what is observed for isotropic $\gamma$ in Refs.~\cite{ShiDCDS2024}.
Note that concave substrate profiles partially compensate for the $\theta > 90^\circ$ contact angle, with final (equilibrium) shapes resembling those obtained with a lower contact angle on a flat substrate (see, e.g., Fig.~\ref{fig::fig4}(a)). Conversely, convex substrates increase the apparent contact angle compared to a planar substrate (see, e.g., Fig.~\ref{fig::fig4}(b)). Such equivalent Winterbottom shape can be computed by considering a wetting parameter $\sigma=\cos\theta_{\rm iso}^\prime$ with an effective angle $\theta_{\rm iso}^\prime=\theta_{\rm iso} - \arcsin(a/(2 R))$, where $R$ is the radius of the substrate (positive for concave substrates and negative for convex substrates) and $a$ is the length of the chord connecting the two contact points. The profiles obtained by this construction are reported as cyan dotted lines in Fig.~\ref{fig::fig4}(a,b) and perfectly overlap with the ones from the PF simulations. 

The same behavior is observed when considering the 3D case, as exemplified in the insets, reporting the final shapes from \ssdw simulation for initially hemispherical particles with the same octahedral $\gamma$-anisotropy of Fig.~\ref{fig::fig3}(a). The apparent contact angle decreases on concave profiles and increases on convex ones, with an effect that becomes more relevant as the particle size grows relative to the substrate radius. This effect holds true for all values of $\sigma$.

\begin{figure}[tb]
    \includegraphics[width=1\linewidth]{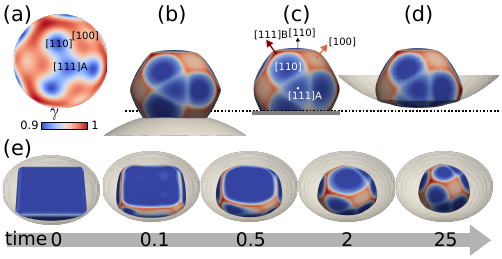}
  \caption{\ssdw simulations for a multifaceted crystal mimicking GaAs. (a) $\gamma$-plot. (b-d) Equilibrium shapes obtained after \ssdw of a hemispherical particle on top of a convex (b), planar (c), and concave (d) spherical substrate. (e) Evolution sequence of the \ssdw of an initially squared GaAs patch filling the bottom of a spherical cap substrate up to $\approx26\%$ of its radius.}
  \label{fig::fig5}
\end{figure}

Thanks to the flexible parametrization of $\gamma$ by eq.~\eqref{eq::msgammma}, it is straightforward to extend the previous analysis to multifaceted crystalline shapes mimicking real materials. In Fig.~\ref{fig::fig5}(a), we consider energy minima ($\hat{\mathbf{m}}_i$) along the $\left\langle100\right\rangle$, $\left\langle110\right\rangle$ and $\left\langle111\right\rangle$ directions and scale them to mimic the anisotropy of GaAs as from Ref.~\cite{MollPRB1996}. In particular, we consider intermediate As chemical potential ($\mu_{\rm As}\approx-0.4$eV) and set $\alpha_{001}=-0.04$, $\alpha_{110}=-0.09$ and $\alpha_{111A}=-0.08$ ($w=18$ for all minima), while \{111\}B facets are omitted due to their higher energy for such conditions. In Fig.~\ref{fig::fig5}(b-d) we report the equilibrium shape obtained as the final state of \ssdw simulations for $\sigma=\cos 120^\circ$, starting from a hemispherical GaAs particle of radius $r$, with vertical axis along the [110] direction, on a convex spherical cap (Fig.~\ref{fig::fig5}(b)), a flat substrate (Fig.~\ref{fig::fig5}(c)) and a concave spherical cap (Fig.~\ref{fig::fig5}(d)). 
In Fig.~\ref{fig::fig5}(e), we report the \ssdw evolution sequence for the same case of Fig.~\ref{fig::fig5}(d) but starting from a rectangular GaAs patch. The final shapes in both these cases converge to the same equilibrium profile.

\begin{figure}[tb]
    \includegraphics[width=1
    \linewidth]{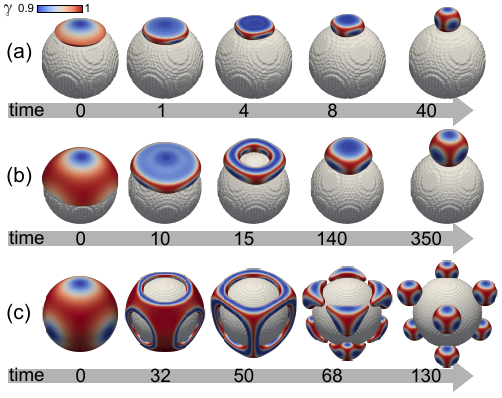}
  \caption{\ssdw simulations of a thin circular film patch with cubic $\gamma$ anisotropy as from Fig.~\ref{fig::fig3}(f) on top of a spherical substrate, for different extent of surface coverage (a) $\approx 9\%$ (b) $\approx 33\%$ and (c) $100\%$. The film thickness-to-substrate radius ratio is $\approx0.09$ (film thickness equal to $2\epsilon$). 
  }\label{fig::fig6}
\end{figure}

A case reminiscent of the \ssdw experiments on oxide nanospheres reported in Ref.~\cite{KrupinskiNANOT2015}, is shown in Fig.~\ref{fig::fig6}. A spherical substrate is considered, and a thin, circular patch of a crystalline film with the cubic $\gamma$ anisotropy of Fig.~\ref{fig::fig3}(f) and contact parameter $\sigma=\cos{120^\circ}$ is deposited on top. In Fig.~\ref{fig::fig6}(a), only a small fraction of the sphere is covered, and the layer retracts from its rims up to reaching the equilibrium shape. By enlarging the film patch size as in Fig.~\ref{fig::fig6}(b), the accumulation of the material at the edges is more relevant, and a different qualitative evolution is obtained. The material tends to be first removed from the patch center, thus leading to an intermediate toroidal shape resembling the film breakup in Fig.~\ref{fig::fig3}(f) at $t=12$. At later stages, however, the shape collapses again into a single particle at the sphere summit. Finally, in Fig.~\ref{fig::fig6}(c), the limiting situation where the film entirely wraps the substrate sphere, as in a core-shell structure, is analyzed. Such initial configuration is stable for isotropic $\gamma$ as due to the uniform chemical potential owing to the constant curvature. On the contrary, for anisotropic $\gamma$, it is unstable as the material flows from the regions aligned along the $\gamma$ minima (maxima of $\mu$) toward the edges (minima of $\mu$), driving the shell faceting. For sufficiently thin shells, this process leads to the opening of holes. A cage-like topology inscribing the substrate sphere forms as the material accumulates at the edges of the cube. Eventually, it breaks into crystalline subunits at the cube corners, each one reaching the equilibrium shape. Different pathways and final arrangements could also result from controlling the relative substrate radius for the configuration in Fig.~\ref{fig::fig6} and, in general, the film thickness as well as its morphology. For example, if the total extension of the film is increased we can observe the same toroidal shape but with the formation of a subunit in the center (see Fig.~S1 in Supplementary Material). Also the shape of the film can change leading to a different arrangement of the crystalline subunits.

\subsection{Dewetting dynamics on surfaces with variable curvature}\label{sec::results::variablek}

\begin{figure*}[tb!]
    \includegraphics[width=1\linewidth]{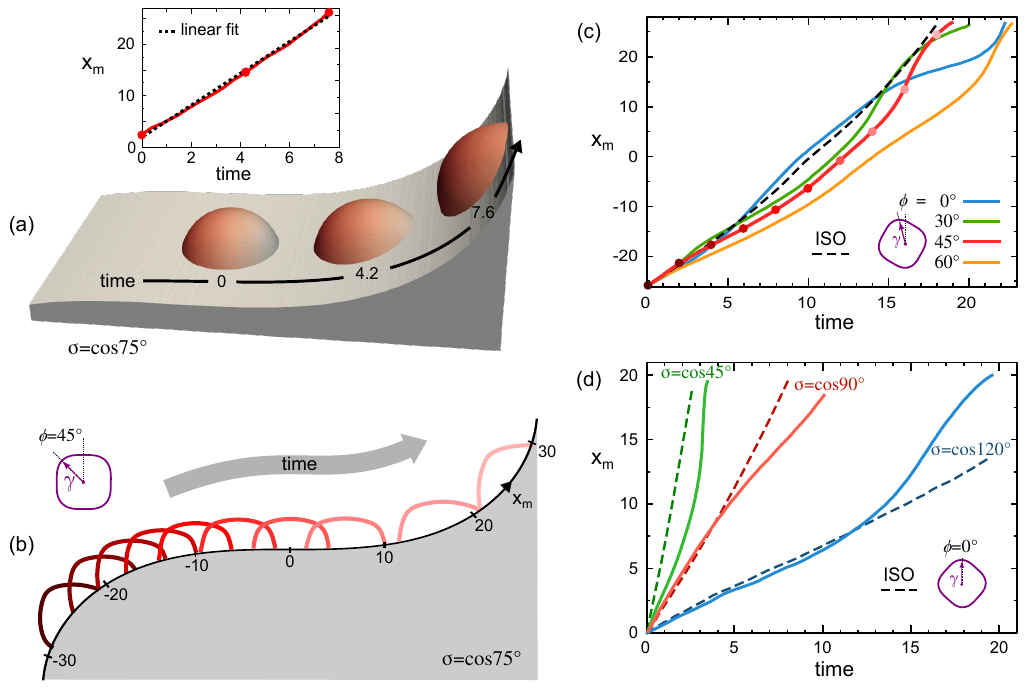}
  \caption{Particle migration by \ssdw along a clothoid substrate. (a) Representative frames of the \ssdw dynamics of an initially hemispherical particle with isotropic $\gamma$ on a 3D clothoid substrate for $\sigma=\cos{75^\circ}$. The inset plot reports the particle displacement as $x_{\rm m}(t)$ (the dots correspond to the three reported frames). (b) Time-lapse sequence of the \ssdw dynamics of an initially hemicircular particle with the four-fold anisotropic $\gamma$ from eq.~\eqref{eq::cosgamma} with $\phi=45^\circ$, on a 2D clothoid substrate for $\sigma=\cos{75^\circ}$. Reported profiles are equally spaced in time. (c,d) $x_{\rm m}(t)$ curve for both isotropic $\gamma$ (dashed lines) and four-fold anisotropic $\gamma$ (solid lines) for (c) different orientation angle $\phi$ and $\sigma=\cos{75^\circ}$ (the case in panel (b) is traced by the red curve, with dots corresponding to each shown profile), and (d) different contact parameter $\sigma=\cos \theta_{\rm iso}$ and $\phi=0^\circ$. All $x_{\rm m}$ values are reported in $\epsilon$ units.}\label{fig::fig7}
\end{figure*}

We now inspect the \ssdw behavior of a (small) particle on top of a substrate with variable curvature. As a first case, in Fig.~\ref{fig::fig7}(a) we consider a substrate set by the 3D extrusion of a clothoid curve. This is a curve with constant curvature gradient along one direction and constant height in the other (see definition in section S1 of Supplementary Material). The dynamics of a hemispherical particle with isotropic $\gamma$ and wetting parameter $\sigma=\cos{75^\circ}$ is reported. Such particle moves along the direction of the curvature gradient, from regions of lower curvature to regions of higher curvature, while retaining a quasi-equilibrium cap shape satisfying the contact BC at all points. This dynamics has been recently explained in Ref.~\cite{ZhaoAM2024}. In particular, such a study showed how, for isotropic $\gamma$, the migration occurs with a velocity linearly dependent on the gradient of the substrate curvature, which in the present case is constant. To check this in our simulation, we tracked the position $x_{\rm m}$ of the center of the contact footprint. The resulting plot, reported as inset in Fig.~\ref{fig::fig7}(a), clearly shows a linear relationship in which the slope corresponds to the constant migration velocity in agreement with the aforementioned theory.

A more complex scenario is found when considering anisotropic $\gamma$ as depending on the relative alignment of facets and contact angles at different positions along the substrate profile. To highlight this relationship, we consider a 2D clothoid substrate and analyze the effects of different $\gamma$ anisotropy and contact angle compared to the isotropic behavior. In Fig.~\ref{fig::fig7}(b), we illustrate the migration along the curvature gradient of a square-faceted particle. Since the timescale of migration is much slower than that of local diffusion, the evolution proceeds as a series of quasi-equilibrium configurations. As the particle moves along the curve, the relative orientation between its facets and the contact direction changes, which in turn leads to a variation in the particle shape. Furthermore, the migration velocity varies significantly along the path, with slower movement before the inflection point, followed by a rapid acceleration afterward. For a more quantitative analysis, we track the particle position by its midpoint $x_{\rm m}$ along the substrate curve; see red curve in Fig.~\ref{fig::fig7}(c). The corresponding $x_{\rm m}(t)$ curve for the case of isotropic $\gamma$ (black dashed line) along with those obtained by rotating the anisotropy function by different angle $\phi$ are included. Clear deviations from the linear behavior of the isotropic reference are found, which also depend on the specific anisotropy. In the case of Fig.~\ref{fig::fig7}(b) ($\phi=45^\circ$) the migration rate is approximately the one of the isotropic case only in the plateau region. Before the inflection this rate is lower and it increases afterwards, with a final deceleration in the last sampled times. Similar trends are found for other values of $\phi$ (see also Fig.~S2 of Supplementary Material) such as $\phi=30^\circ$, showing an even stronger deceleration at latest stages, or $\phi=60^\circ$, characterized by a slower overall velocity. For $\phi=0^\circ$ the trend is instead almost inverted with super-linear behavior in the first stages followed by a significant slow-down after the inflection point and an acceleration in the final stages. These differences result from the relative orientation of crystal facets with respect to the substrate slope at the different positions along the curve.

As discussed in Ref.~\cite{JiangPRM2018}, in the case of isotropic $\gamma$, the migration velocity decreases when increasing the contact angle $\theta_{\rm iso}$ due to the corresponding reduction of the contact area with the substrate. The same trend is achieved in our PF simulation with isotropic $\gamma$ as shown by the slopes of the dashed lines in Fig.~\ref{fig::fig7}(d) (a detailed comparison with the analytic prediction from Ref.~\cite{ZhaoAM2024} is reported in the Fig.~S2(c) of Supplementary Material). Importantly, we obtain that this trend still holds in the presence of surface-energy anisotropy, even if additional effects and changes of the migration velocity emerge. This is illustrated by solid lines in Fig.~\ref{fig::fig7}(d): although $x_{\rm m}(t)$ does not feature a constant slope owing to the varying morphology (see again Fig.~\ref{fig::fig7}(b)), the overall trend well corresponds to the isotropic case.

\begin{figure}[tb]
    \includegraphics[width=\columnwidth]{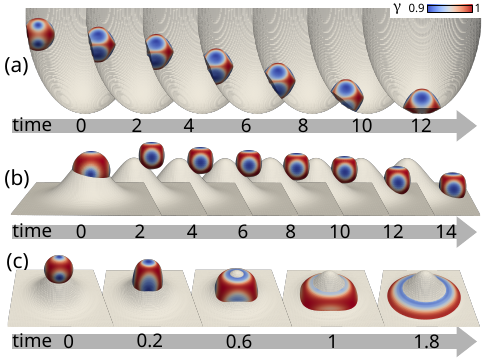}
  \caption{Migration by \ssdw of a 3D crystalline particle on curved substrates. (a) Trajectory of a particle with the octahedral $\gamma$ anisotropy of Fig.~\ref{fig::fig3}(a) moving on a $2:3:4$ ellipsoidal cavity with $\sigma=\cos120^\circ$. (b,c) Trajectory of a particle with the cubic $\gamma$ anisotropy of Fig.~\ref{fig::fig3}(f) moving on a gaussian-shaped substrate with (b) $\sigma=\cos120^\circ$ and (c) $\sigma=\cos45^\circ$. The particle is initialized as a spherical cap of diameter equal to the peak height. In (c) it is centered at the summit of the gaussian while in (b) the initial particle is offset to the right by 20\% of its radius to trigger migration.}\label{fig::fig8}
\end{figure}

As a more general case, in Fig.~\ref{fig::fig8}, we report the motion of faceted particles on arbitrary 3D curves. In Fig.~\ref{fig::fig8}(a) an ellipsoidal substrate cavity is considered and the migration of a crystal particle with the octahedral $\gamma$ anisotropy of Fig.~\ref{fig::fig3}(a) is shown. As expected, the particle migrates from the initial region of low curvature toward the final position of maximum curvature. The frames are reported at equal time intervals, showing that the velocity of the particle increases in the last stages, where the curvature gradient grows larger. Another example is reported in Fig.~\ref{fig::fig8}(b,c), where a small particle with cubic anisotropy (as in Fig.~\ref{fig::fig3}(f)) is deposited on top of a Gaussian peak. In Fig.~\ref{fig::fig8}(b), a wetting parameter $\sigma=\cos{120^\circ}$ is set, and a faceted particle is formed on top of the peak. However, that is an unstable equilibrium position. A small deviation in the particle centering with respect to the peak of the Gaussian profile would initiate its migration on one side. Here, this effect has been triggered by a slight shift in the initial position of the island. A completely different evolution path is observed for $\sigma=\cos{45^\circ}$ as shown in Fig.~\ref{fig::fig8}(c). The particle spreads around the peak sides, and due to the variation of the substrate curvature, it is dragged downward, resulting in a ring shape. As the internal radius aligns with the peak diameter and expands as the crystal front descends along it, the crystalline ring gradually becomes thinner. We can also notice how the time scale of the simulations in Fig.~\ref{fig::fig8}(b) and Fig.~\ref{fig::fig8}(c) are very different, with the latter being much faster than the former, even though the two configurations are comparable in size, in agreement with what observed in Fig.~\ref{fig::fig7}(c).

\subsection{Dewetting on pit-patterned substrates}\label{sec::results::pits}

We finally investigate the dynamics of dewetting on pit-patterned substrates as a potential platform to achieve complex structures and particle distributions controlled by the chosen topography. Regular arrays of pits have already been discussed in a few literature studies \cite{OhSMALL2009,WangJMSME2011,GiermannJAP2011,MukherjeeSM2008} to localize the dewetted particles at targeted positions. Indeed, in the presence of surface modulations, material flow is triggered by capillarity forces from convex to concave regions. For thin-enough films, dewetting starts simultaneously at each pit site, which may then realize a higher uniformity throughout the patterned substrate besides controlling the final shapes and connectivity among dewetted structures. Still, detailed modelling and prediction of this process need to account for the three-dimensionality of the evolving geometries and energy contribution of all interfaces, aspects now accessible with the proposed PF framework.

\begin{figure}[tb!]
    \includegraphics[width=\columnwidth]{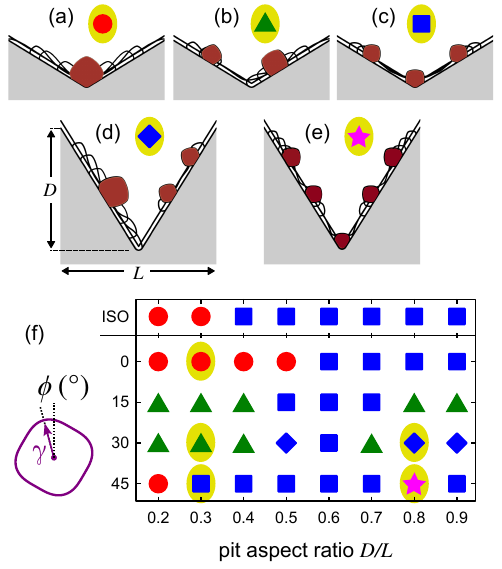}
  \caption{\ssdw of a 2D film with four-fold anisotropy from eq.~\eqref{eq::cosgamma}, deposited conformally onto a triangular pit with fixed $h/L=0.025$ film thickness (here $L=100 \epsilon$) and $\sigma$=$\cos$135$^\circ$. (a-e) Evolution sequences for different final configurations starting from the same film thickness on pits of different depth-to-width aspect ratio: (a-c) $D/L=0.3$, (d,e) $D/L=0.8$, and different orientation $\phi$ of the crystallographic axis relative to the substrate: (a) $\phi=0^\circ$, (b) $\phi=15^\circ$, (c,e) $\phi=45^\circ$, (d) $\phi=30^\circ$. (f) Schematic representation illustrating the different equilibrium configurations (single island at pit center, two islands asymmetrically on the pit sidewalls and three islands, one at the center and two symmetrical on the sides) as a function of $D/L$ and $\phi$. The case of isotropic $\gamma$ is also included.}\label{fig::fig9}
\end{figure}

In Fig.~\ref{fig::fig9}, we consider a 2D triangular pit conformally covered by a thin film and present a systematic analysis on how the final shape and the number of islands depend on the pit depth-to-width aspect ratio $D/L$ upon \ssdwnospace. Either isotropic $\gamma$ or the four-fold anisotropic function of Fig.~\ref{fig::fig2} are considered, with the latter inspected for different orientations $\phi$ of the crystallographic axis. Specifically, we change the pit depth $D$ while keeping the same width $L$ with $D/L \in [0.2, 0.8]$, which corresponds to an angular opening of the pit ranging from approximately $135^\circ$ to $60^\circ$. Reported results refer to a contact parameter $\sigma = \cos{135^\circ}$, but similar trends have also been obtained for different conditions. In all cases, the film thickness is set to $h=0.025 L$. Reducing the $h/L$ ratio favors the film breakup into more sub-units under identical conditions. 
As expected, the initial continuous film evolves under the action of the capillarity driving force, inducing a material flow toward the pit bottom, which in turn causes a progressive thinning at the upper rims, eventually leading to film breakup. After this initial splitting, different outcomes are possible depending on the interplay between the pit geometry and film anisotropy. This process can either lead to a single island at the pit bottom or multiple sub-units in different positions, as shown by the representative cases in Fig.~\ref{fig::fig9}(a-e). An overview of different configurations obtained by varying $\sigma$, $D/L$ and $\phi$ is reported in Figs.~S3 and S4 of Supplementary Material.

A comprehensive view of the different numbers and arrangements of islands for various $\gamma$ and $D/L$ is shown in Fig.~\ref{fig::fig9}(f). In the case of isotropic $\gamma$, increasing the $D/L$ ratio beyond a certain threshold leads to the transition from one to three crystalline particles. This can be easily explained by considering the increment of the film length. While this effect is expected to play a role also when $\gamma$ is anisotropic, the actual trends in the Fig.~\ref{fig::fig9}(f) are not as straightforward. In the case of $\phi=0^\circ$, the same transition from one to three islands is observed. However, the $D/L$ threshold shifts to higher values ($D/L=0.5$) due to the stabilization of the film when the pit sidewalls are nearly parallel to the film top facet. The opposite happens when considering a rotation of $\phi=45^\circ$ of the definition of $\gamma$. In such a case, indeed, the crystal facets always cut the pit sidewalls at oblique angles, thus destabilizing the film. In this case, the transition from one to three islands is obtained at low $D/L$ ratios. Film splitting thus depend on the relative orientation of the crystal facets to the pit sides. Notably, the different onset of the film breakup due to the relative alignment of the crystal facets with the pit sidewalls also determines the material redistribution between the island at the pit bottom and the ones on the sidewalls. For instance, this is responsible for the behavior observed for $\phi=45^\circ$ and $D/L=0.8$ where the formation of a smaller island at the bottom leaves sufficient material along the sidewalls to observe a second breakup and a final five-island arrangement; see Fig.~\ref{fig::fig9}(e). However, this effect comes from a subtle balance between capillarity and edge retraction dynamics, which is lost when further increasing the $D/L$ aspect ratio where more material moves toward the bottom. Similarly, a three-island configuration for the the same $D/L$ ratio is also recovered for stronger wetting conditions, i.e., by setting $\sigma=\cos{120^\circ}$ or a lower contact angle (see Figs.~S3 and S4 of Supplementary Material).

For $\phi$ rotation angles in between the two limiting cases discussed above, the asymmetrical orientation of the facets with respect to the pit sidewalls returns a different dynamics of edge retraction on the two pit sides. As shown for both the cases with $\phi=15^\circ$ and $30^\circ$ in Fig.~\ref{fig::fig9}(f), this destabilizes the film favoring splitting even for the lowest $D/L$. In both cases, the final configuration on the shallowest pits up to $D/L=0.4$ consists of two asymmetric islands, as exemplified in Fig.~\ref{fig::fig9}(b). In particular, the smallest island forms along the pit sidewall featuring the worst alignment with the crystal facets. Also, it forms at a higher quota as the splitting is accelerated on such side. The island on the other sidewall, thanks to the better facet alignment, acquires more material and hence grows larger and closer to the pit bottom. This uneven material distribution results in asymmetric additional breakups for higher $D/L$ ratio; see, e.g., Fig.~\ref{fig::fig9}(d). However, this situation is found only for $\phi=30^\circ$ and for some $D/L$.  For intermediate $D/L$ values ($\approx0.6$), instead, a larger amount of material is found to accumulate at the pit bottom so that a third island is formed there, giving a final configuration similar to the one of Fig.~\ref{fig::fig9}(c) although asymmetric.

\begin{figure*}[tb!]
    \includegraphics[width=1 \linewidth]{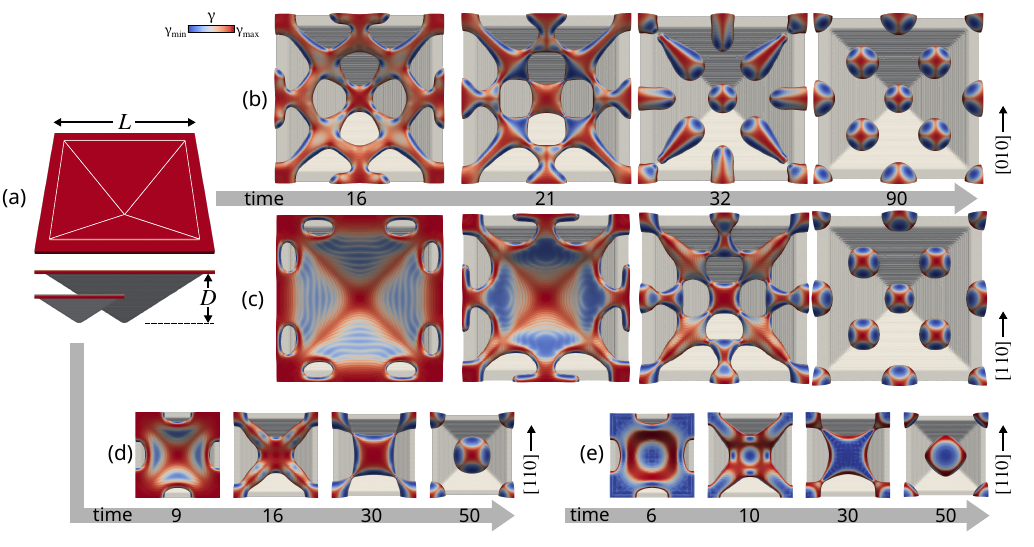}
  \caption{3D \ssdw simulations of a thin film deposited conformally on a square-based pyramidal pit with depth-to-width aspect ratio $\approx0.35$ and two sizes, one double the other, as shown in (a). The initial film thickness is set equal to $D/12$ of the smallest pit. (b-e) Evolution of the film morphology in top view at representative times. In (b-d) the \{111\} octahedral $\gamma$ anisotropy of Fig.~\ref{fig::fig3}(a) is considered while in (e) the \{100\} cubic one of Fig.~\ref{fig::fig3}(f) is used. The relative orientation of the crystallographic axis with respect to the pit sides is indicated for each case and the film is colored according to the local values of $\gamma$.}\label{fig::fig10}
\end{figure*}

While the previous discussion focused on the final state, the interplay between anisotropy and pit geometry also affects the full kinetic pathway of the \ssdw process. This is illustrated in Fig.~\ref{fig::fig10} by considering a square-based pyramidal pit bounded by a small flat region ($\approx 18\%$ of the lateral size $L$) separating each pit from the other in a hypothetical periodic array. The geometry is shown in Fig.~\ref{fig::fig10}(a), also illustrating two pit sizes sharing the same depth-to-width aspect ratio ($D/L\approx0.35$) and film thickness (set equal to $D/12$ of the smallest pit) used in the simulations discussed below. In Fig.~\ref{fig::fig10}(b,c), we report and compare the simulations considering the octahedral anisotropy used in Fig.~\ref{fig::fig3}(a) oriented with [001] vertical axis and rotated about the vertical axes such to have stable film facets aligned with the edges (Fig.~\ref{fig::fig10}(b)) or the sidewalls (Fig.~\ref{fig::fig10}(c)) of the pit. The two evolutions closely resemble each other, featuring similar morphologies forming, however, at different times. Consistently, a slower dynamic is observed for film facets aligned with pit sidewalls. The film breaks as the result of material flux triggered by capillarity, as in all the cases discussed above. This is found first to occur close to the top corners. Further breakups occur at the center of the sidewalls, and the evolution ends with the formation of individual faceted particles within and outside the pit.
The size of the pits affects the outcome in terms of morphology during the dewetting process as well as the final number of isolated islands. This is shown in Fig.~\ref{fig::fig10}(d) where the same settings of Fig.~\ref{fig::fig10}(c) is considered but for a pit with half the size (see Fig.~\ref{fig::fig10}(a)). In this case, holes open only at the center of pit edges, then material retracts along the pit diagonals and finally breaks into a large crystal unit in the pit center and small ones at the pit upper corners. 
Last, the simulation of Fig.~\ref{fig::fig10}(e) considers the cubic $\gamma$ anisotropy as in Fig.~\ref{fig::fig3}(f) (for the smallest considered pit size). The dynamics looks similar to the one in Fig.~\ref{fig::fig10}(d) despite the different $\gamma$ anisotropy, thus indicating how the capillarity effects are here playing the major role with respect to the local re-faceting. Also in this case, the different alignment of the crystal facets with respect to the dewetting front affects the relative velocity of the process, although with less impact than on the larger pit discussed above. 

Finally, it is worth pointing out that the pit morphology in real experiments would generally correspond to smoother profiles and sidewalls rather than straight planes as the idealized reversed-pyramids of the previous simulations. For this reason, local curvature gradients are generally present. According to the discussion of Fig.~\ref{fig::fig8}, they could induce migration of the crystalline particles. In particular, if multiple particles are formed at the sidewalls of a smooth pit (e.g., parabolic or Gaussian), they will slowly move toward the higher curvature bottom so that, in the long run, they eventually coalesce back again into a single particle. While this effect could be minor as involving long timescales, especially for almost flat sidewalls and almost sharp edges, the \ssdw dynamics generating the particles may still be affected by variation in the actual positions of the dewetted particles. Still, the proposed analysis and discussion outlines the key elements to be considered to acquire an in-depth control of \ssdw on non-flat substrates, and strategies driving and complementing experimental campaigns are envisaged.

\section{Conclusion} \label{sec:conclusions}
A PF model describing the dynamics of \ssdw of crystalline, faceted particles and thin films on arbitrary substrate morphologies has been developed. The implicit definition of the substrate profile via the $\psi$ order parameter allows for varying its topography without reshaping the integration domain, which can be simply chosen as a cuboid box. Moreover, it naturally embeds all the geometrical information needed for the proper tracking of the film-substrate contact points without the need to explicitly describe their motion as in sharp interface approaches. These advantages make the present approach extremely flexible and seamlessly applicable to two- or three-dimensional problems. Although the present study limits to fully-convex $\gamma$ anisotropy, the same framework could be identically used also in the case of strong anisotropy, provided that an appropriate corner regularization is implemented. The generalization of the present approach to include this regime will be the object of future work.

The resulting description of \ssdw for the late evolution stages is shown to be consistent with the theoretical equilibrium shapes predicted by the Winterbottom construction. Moreover, the one-to-one match of the entire evolution sequences with a conventional PF model on planar geometries confirms the model reliability in predicting the actual kinetic pathway toward equilibrium. A set of novel simulations of \ssdw on nonplanar substrates have been then delivered, addressing theoretical aspects as well as scenarios mimicking experimental complexity.

The several cases inspected in this work highlight in particular the effects of using a non-planar substrate geometry in \ssdw for crystalline thin films, including the change of the apparent contact angles on spherical supports, the dynamics of particle migration induced by non-constant substrate curvatures, and the triggering of complex topological changes stemming from the combined effects of wetting and capillarity. We thus suggest approaches for tailoring a wide variety of configurations, ranging from the ordered crystalline particles to complex (metastable) morphologies and topologies, by the careful choice of substrate geometry and crystallographic orientation of the thin film. These strategies can be exploited as an alternative/addition to state-of-the-art thin-film patterning approaches.
\newpage
\section*{Acknowledgments}
M.S. acknowledges support from the Deutsche Forschungsgemeinschaft (DFG, German Research Foundation) within FOR3013, project number 417223351.
E.R. acknowledges support from the doctoral project DurAMat under the European Union - Marie Sk\l{}odowska-Curie Actions Doctoral Networks (ITN) Call H2020-MSCA-DN-2022, grant no. 101119767.

\bibliography{biblio}

\noindent 

\clearpage

\makeatletter

\def\maketitle{
\@author@finish
\title@column\titleblock@produce
\suppressfloats[t]}
\makeatother

\setcounter{equation}{0}
\setcounter{figure}{0}
\setcounter{table}{0}
\setcounter{page}{1}
\setcounter{section}{0}
\makeatletter
\long\def\MaketitleBox{%
  \resetTitleCounters
  \def\baselinestretch{1}%
  \begin{center}%
   \def\baselinestretch{1}%
    \Large\@title\par\vskip18pt
    \normalsize\elsauthors\par\vskip10pt
    \footnotesize\itshape\elsaddress\par\vskip36pt
    \end{center}
  }
\makeatother

\title{\hrule\vspace{1cm} SUPPLEMENTARY MATERIAL \\\vspace{1cm}\hrule\vskip0pt\vspace{1.cm} Phase-field Modelling of Anisotropic Solid-State Dewetting on Patterned Substrates}

\maketitle
\onecolumngrid

\renewcommand{\thefigure}{S\arabic{figure}}
\renewcommand{\thesection}{S\arabic{section}}

\section{Parametric definition of the clothoid curve}
The clothoid curve $\mathcal{C}(s)\equiv (x(s),y(s))$ is characterized by a linear change in the curvature with its arc length $s$. It can be described by the following parametrization: 
\begin{equation*}
x(s) = \int_0^s \cos\left(\frac{\kappa^\prime u^2}{2}\right) \, du, \qquad 
y(s) = \int_0^s \sin\left(\frac{\kappa^\prime u^2}{2}\right) \, du,
\end{equation*}
where \(\kappa^\prime\) is the curvature gradient. In the simulations reported in this work illustrating dewetting on a substrate with this shape (Fig.~$7$ of main manuscript and Fig.~\ref{fig::S2}), we consider $\kappa^\prime=0.0028$. Its extension to 3D is obtained by considering $\mathcal{C}(s)\equiv (x(s),y(s),z(s))$ with $z(s)=z$ independent of $s$ (an extrusion of the 2D profile). 

\vspace{1cm}
\section{Additional simulations of dewetting on nonplanar substrates}

\begin{figure}[ht!]
    \centering
    \includegraphics[width=0.75\textwidth]{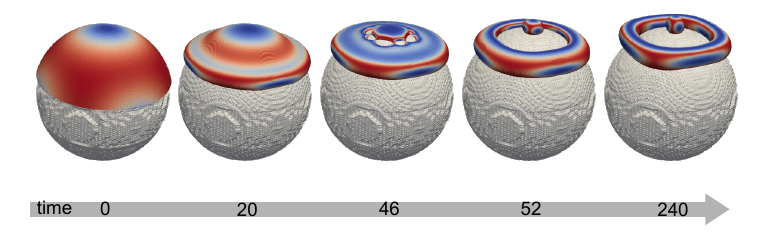}
    \caption{Simulation of solid-state dewetting of a circular patch of thin film with cubic $\gamma$ anisotropy on top of a spherical substrate. Parameters are the same of Fig.~6(b) of the main manuscript but for an increment of a factor 1.8 in the sphere diameter. Similarly to Fig.~6(b), the retraction of the film patch results into a toroidal configuration but, given the larger size, enough material is left at the sphere summit to produce an island. Eventually, both objects merge into a single larger island, converging to the same equilibrium state of Fig.~6(b).}
    \label{fig::S1}
\end{figure}

\begin{figure}[t!]
    \centering
    \includegraphics[width=\textwidth]{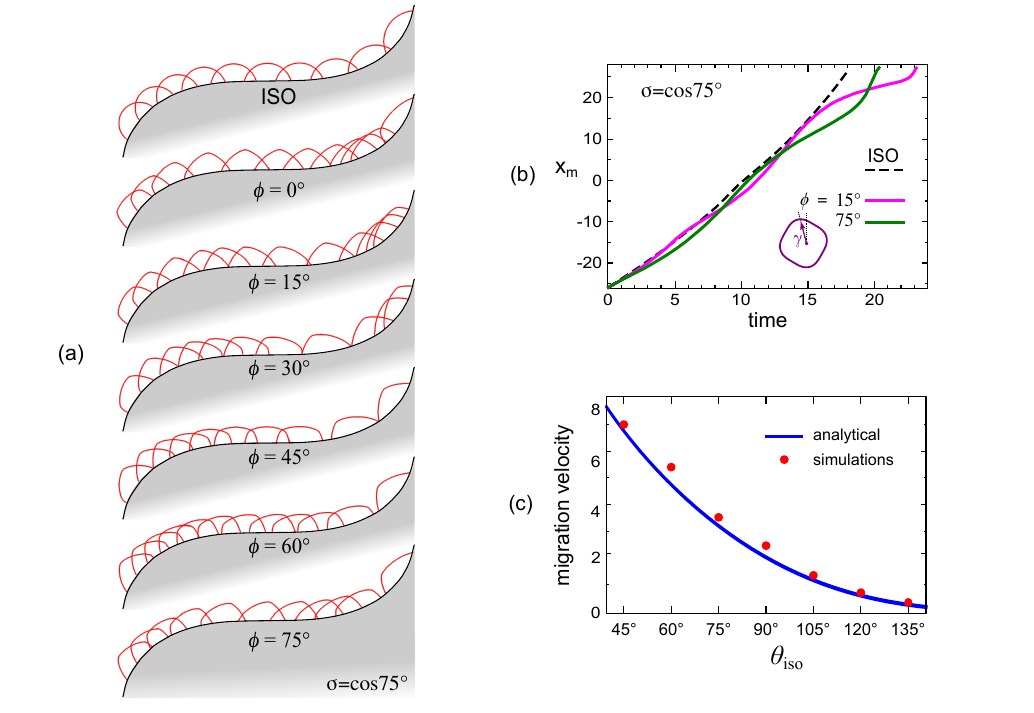}
    \caption{Particle migration along a clothoid curve. (a) Time-lapse sequences for the motion of an initially hemicircular particle with four-fold anisotropic $\gamma$ for different rotation angles $\phi$ corresponding to the cases shown in Fig.~$7$(b) and (c) of the main manuscript. (b) $x_{\rm m}(t)$ for the cases with $\phi=15^\circ$ and $75^\circ$, along with the one for isotropic $\gamma$ (dashed) for comparison. (c) Plot of the migration velocity of a 2D isotropic particle as a function of the contact angle $\theta_{\rm iso}$. The velocities are obtained as the slope of a linear fit of the $x_{\rm m}(t)$ around the clothoid inflection point. The simulation results (dots) are compared with the analytical behavior expected from the study in \href{https://doi.org/10.1016/j.actamat.2024.120407}{Q. Zhao et al. Acta Materialia 281 (2024) 120407}.}
    \label{fig::S2}
\end{figure}

\begin{figure}[ht!]
    \centering
     \includegraphics[width=\textwidth]{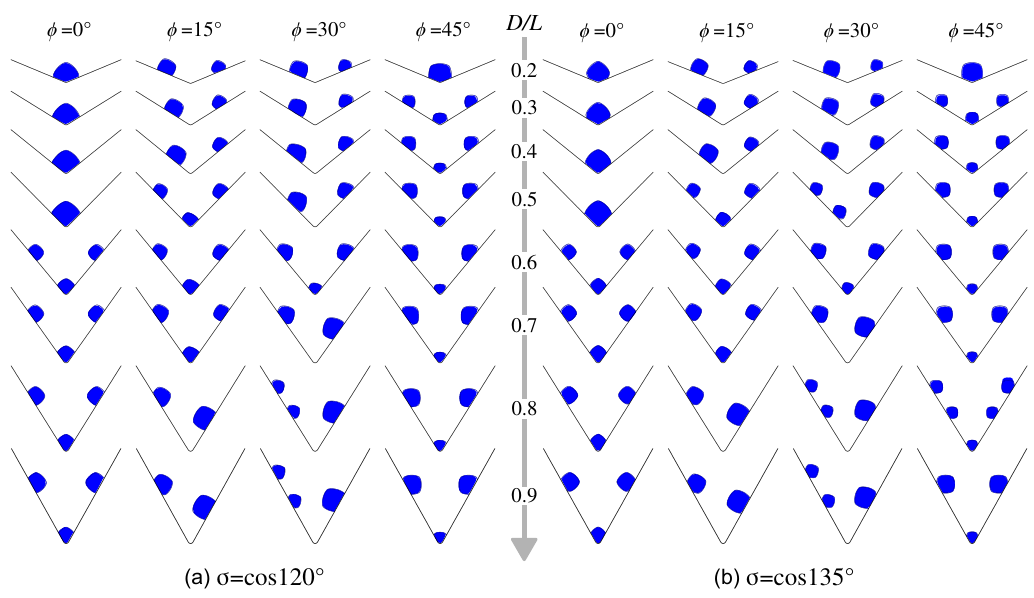}
    \caption{Final configurations from simulations of solid state dewetting of a 2D film with four-fold anisotropy, deposited conformally onto a triangular pit with fixed $h/L=0.025$ film thickness and contact parameter (a) $\sigma=\cos120^\circ$ and (b) $\sigma=\cos135^\circ$ (same as Fig.~9 of main manuscript). The reported cases are ordered as a function of depth-to-width $D/L$ pit aspect ratio and rotation angle $\phi$ of the anisotropic surface energy density $\gamma$.}
    \label{fig::S3}
\end{figure}

\begin{figure}[ht!]
    \centering
    \includegraphics[width=\textwidth]{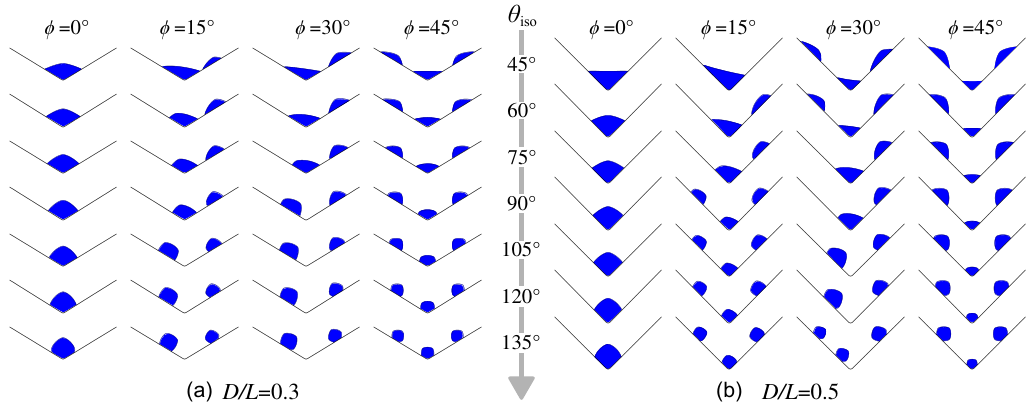}
    \caption{Final configurations from simulations of solid state dewetting of a 2D film with four-fold anisotropy, deposited conformally onto a triangular pit with depth-to-width $D/L$ aspect ratio of (a) $0.3$ and (b) $0.5$. The reported cases are ordered as a function of contact parameter $\theta_{\rm iso}$ and rotation angle $\phi$ of the anisotropic surface energy density $\gamma$.}
    \label{fig::S4}
\end{figure}

\end{document}